\pdfoutput=1
\documentclass[10pt,logo,copyright]{nvidiatechreport}
\usepackage[authoryear,sort&compress,round]{natbib}

\usepackage[utf8]{inputenc} 
\usepackage[T1]{fontenc}    

\usepackage{parskip}        
\usepackage{url}            
\usepackage{hyperref}       
\usepackage{booktabs}       
\usepackage{amsfonts}       
\usepackage{nicefrac}       
\usepackage{microtype}      
\usepackage{xcolor}         
\usepackage[dvipsnames]{xcolor} 
\usepackage{graphicx}
\usepackage{animate}        
\usepackage{subcaption}
\usepackage{tabularx}
\usepackage{makecell}
\usepackage{adjustbox}
\usepackage{setspace}
\newcolumntype{M}[1]{>{\centering\arraybackslash}m{#1}}
\usepackage{float}
\usepackage{tikz}
\usetikzlibrary{positioning,shapes,arrows}
\usepackage{amsmath,amsfonts,bm, bbm,leftindex}
\usepackage{multirow}
\usepackage{comment}
\usepackage{gensymb}
\usepackage{lipsum}
\usetikzlibrary{arrows.meta, positioning, fit}
\usepackage[para]{threeparttable}
\usepackage{tikz}
\usetikzlibrary{tikzmark}

\newcommand{\captiona}{{\em (a)}}
\newcommand{\captionb}{{\em (b)}}

\def\eqref#1{equation~\ref{#1}}

\def\1{\bm{1}}

\DeclareMathAlphabet{\mathsfit}{\encodingdefault}{\sfdefault}{m}{sl}
\SetMathAlphabet{\mathsfit}{bold}{\encodingdefault}{\sfdefault}{bx}{n}

\makeatletter
\let\save@mathaccent\mathaccent
\newcommand*\if@single[3]{%
  \setbox0\hbox{${\mathaccent"0362{#1}}^H$}%
  \setbox2\hbox{${\mathaccent"0362{\kern0pt#1}}^H$}%
  \ifdim\ht0=\ht2 #3\else #2\fi
  }
\newcommand*\rel@kern[1]{\kern#1\dimexpr\macc@kerna}
\newcommand*\widebar[1]{\@ifnextchar^{{\wide@bar{#1}{0}}}{\wide@bar{#1}{1}}}
\newcommand*\wide@bar[2]{\if@single{#1}{\wide@bar@{#1}{#2}{1}}{\wide@bar@{#1}{#2}{2}}}
\newcommand*\wide@bar@[3]{%
  \begingroup
  \def\mathaccent##1##2{%
    \let\mathaccent\save@mathaccent
    \if#32 \let\macc@nucleus\first@char \fi
    \setbox\z@\hbox{$\macc@style{\macc@nucleus}_{}$}%
    \setbox\tw@\hbox{$\macc@style{\macc@nucleus}{}_{}$}%
    \dimen@\wd\tw@
    \advance\dimen@-\wd\z@
    \divide\dimen@ 3
    \@tempdima\wd\tw@
    \advance\@tempdima-\scriptspace
    \divide\@tempdima 10
    \advance\dimen@-\@tempdima
    \ifdim\dimen@>\z@ \dimen@0pt\fi
    \rel@kern{0.6}\kern-\dimen@
    \if#31
      \overline{\rel@kern{-0.6}\kern\dimen@\macc@nucleus\rel@kern{0.4}\kern\dimen@}%
      \advance\dimen@0.4\dimexpr\macc@kerna
      \let\final@kern#2%
      \ifdim\dimen@<\z@ \let\final@kern1\fi
      \if\final@kern1 \kern-\dimen@\fi
    \else
      \overline{\rel@kern{-0.6}\kern\dimen@#1}%
    \fi
  }%
  \macc@depth\@ne
  \let\math@bgroup\@empty \let\math@egroup\macc@set@skewchar
  \mathsurround\z@ \frozen@everymath{\mathgroup\macc@group\relax}%
  \macc@set@skewchar\relax
  \let\mathaccentV\macc@nested@a
  \if#31
    \macc@nested@a\relax111{#1}%
  \else
    \def\gobble@till@marker##1\endmarker{}%
    \futurelet\first@char\gobble@till@marker#1\endmarker
    \ifcat\noexpand\first@char A\else
      \def\first@char{}%
    \fi
    \macc@nested@a\relax111{\first@char}%
  \fi
  \endgroup
}
\makeatother

\definecolor{darkred}{rgb}{0.7, 0.0, 0.0}

\usepackage{pifont}
\newcommand{\cmark}{\ding{51}} 
\newcommand{\xmark}{\ding{55}} 
\newcommand{\pmark}{\textcolor{gray}{\ding{108}}} 
\usepackage{wrapfig}
\renewcommand{\today}{}

\usepackage[nameinlink]{cleveref}
\crefname{equation}{Eq.}{Eqs.}
\crefname{figure}{Fig.}{Figs.}
\crefname{section}{Sec.}{Sec.}
\crefname{appendix}{App.}{App.}
\crefname{table}{Tab.}{Tabs.}
\crefname{algorithm}{Algo}{Algo}
\crefname{thm}{Thm}{Thm}
\Crefname{thm}{Thm}{Thm}
\crefname{prop}{Prop}{Prop}
\usepackage{longtable}
\usepackage{wrapfig}
\usepackage{caption}
\usepackage{makecell}
\usepackage{xurl}
\usepackage{hyperref}
\usepackage{xcolor}
\usepackage{hyperref}
\usepackage{tikz}
\usetikzlibrary{trees} 
\usepackage[edges]{forest}
\usepackage[breakable]{tcolorbox}
\definecolor{nvidiaGreen}{HTML}{9dca63}
\newcommand{\crefnames}[3]{%
  \@for\next:=#1\do{%
    \expandafter\crefname\expandafter{\next}{#2}{#3}%
  }%
}
\usepackage{listings}
\lstdefinestyle{pythonstyle}{language=Python, basicstyle=\ttfamily\small, keywordstyle=\color{blue}, commentstyle=\color{gray}, stringstyle=\color{red}, showstringspaces=false, breaklines=true}
\tcbuselibrary{listings,skins}
\definecolor{polarPanelTitle}{HTML}{3f3f3f}
\definecolor{polarPanelBody}{HTML}{f1f2ff}
\definecolor{polarPanelFrame}{HTML}{4a4a4a}
\newtcblisting{polarlisting}[1]{
    listing only,
    breakable,
    enhanced,
    title={#1},
    colback=polarPanelBody,
    colframe=polarPanelFrame,
    colbacktitle=polarPanelTitle,
    coltitle=white,
    fonttitle=\bfseries,
    boxrule=0.7pt,
    arc=3pt,
    left=7pt,
    right=7pt,
    top=6pt,
    bottom=6pt,
    listing options={
        language=Python,
        basicstyle=\ttfamily\footnotesize,
        keywordstyle=\color{black},
        commentstyle=\color{black!55},
        stringstyle=\color{black!70},
        showstringspaces=false,
        breaklines=true,
        columns=fullflexible,
        keepspaces=true
    }
}

\definecolor{midnightgreen}{rgb}{0.0, 0.29, 0.33}
\definecolor{deepgreen}{HTML}{0aa344}

\newcommand{\nvgreen}[1]{\textcolor{nvidiagreen}{#1}}
\definecolor{deeppurple}{HTML}{7030a0}
\definecolor{deepblue}{HTML}{171d91}
\definecolor{brown}{HTML}{843c0c}
\definecolor{shadered}{HTML}{ffe5e5}
\definecolor{shadegreen}{HTML}{e5f7ed}
\definecolor{msftBlack}{RGB}{0,0,0}
\definecolor{lightred}{RGB}{255,163,163}
\definecolor{deepred}{RGB}{146,0,0}
\newtcolorbox{boxL}{
    fontupper = \color{black},
    rounded corners,
    arc = 6pt,
    colframe = black!50,
    boxrule = 0pt,
    bottomrule = 4.5pt ,
    breakable,
}
\newcommand{\polar}{\textsc{Polar}\xspace}
\newcommand{\prorlagent}{\textsc{ProRL Agent}\xspace}
\title{Polar: Agentic RL on Any Harness at Scale}
\author{Binfeng Xu, Hao Zhang, Shaokun Zhang, Songyang Han, Mingjie Liu, Jian Hu, Shizhe Diao, Zhenghui Jin, Yunheng Zou, Michael Demoret, Jan Kautz, Yi Dong
}

\begin{abstract}

Reinforcement learning for language agents increasingly depends on custom harnesses that manage long-running context, multi-turn tool use and multi-agent orchestration. 
However, porting these harnesses into RL environment interfaces remains difficult and often loses important training signals.
We bridge this gap with \nvgreen{\polar}, a rollout framework for scalable asynchronous RL over arbitrary agent harnesses.
\polar treats the agent harness as a black box: it proxies LLM API calls, records token-level model interactions, and reconstructs token-faithful trajectories for training. Each rollout node efficiently manages runtime prewarming, agent execution, trajectory reconstruction, and evaluation in parallel, exposing asynchronous service endpoints that can be consumed by independent trainers at scale.
This decoupled design makes \polar agnostic to agent harnesses, training infrastructure, and RL algorithms while improving compute utilization for long-running agent workloads.
We validate \polar by training agents on software-engineering tasks with popular coding harnesses.
Using simple GRPO, \polar improves Qwen3.5-4B by 22.6, 4.8, 0.6 and 6.2 points on SWE-Bench Verified with the Codex, Claude Code, Qwen Code and Pi harnesses, respectively.
We further demonstrate \polar for offline data generation over custom harnesses and ablate trajectory reconstruction strategies. \nvgreen{\polar} rewrites its preceding work, 
\nvgreen{\prorlagent}\footnotemark{}
and has been registered as one of \nvgreen{NeMo Gym} environments.
\end{abstract}

\begin{document}

\maketitle
\abscontent
\footnotetext{Code available at \url{https://github.com/NVIDIA-NeMo/ProRL-Agent-Server}\vspace{-5mm}}

\begin{figure}[H]
\centering
\includegraphics[width=0.82\textwidth]{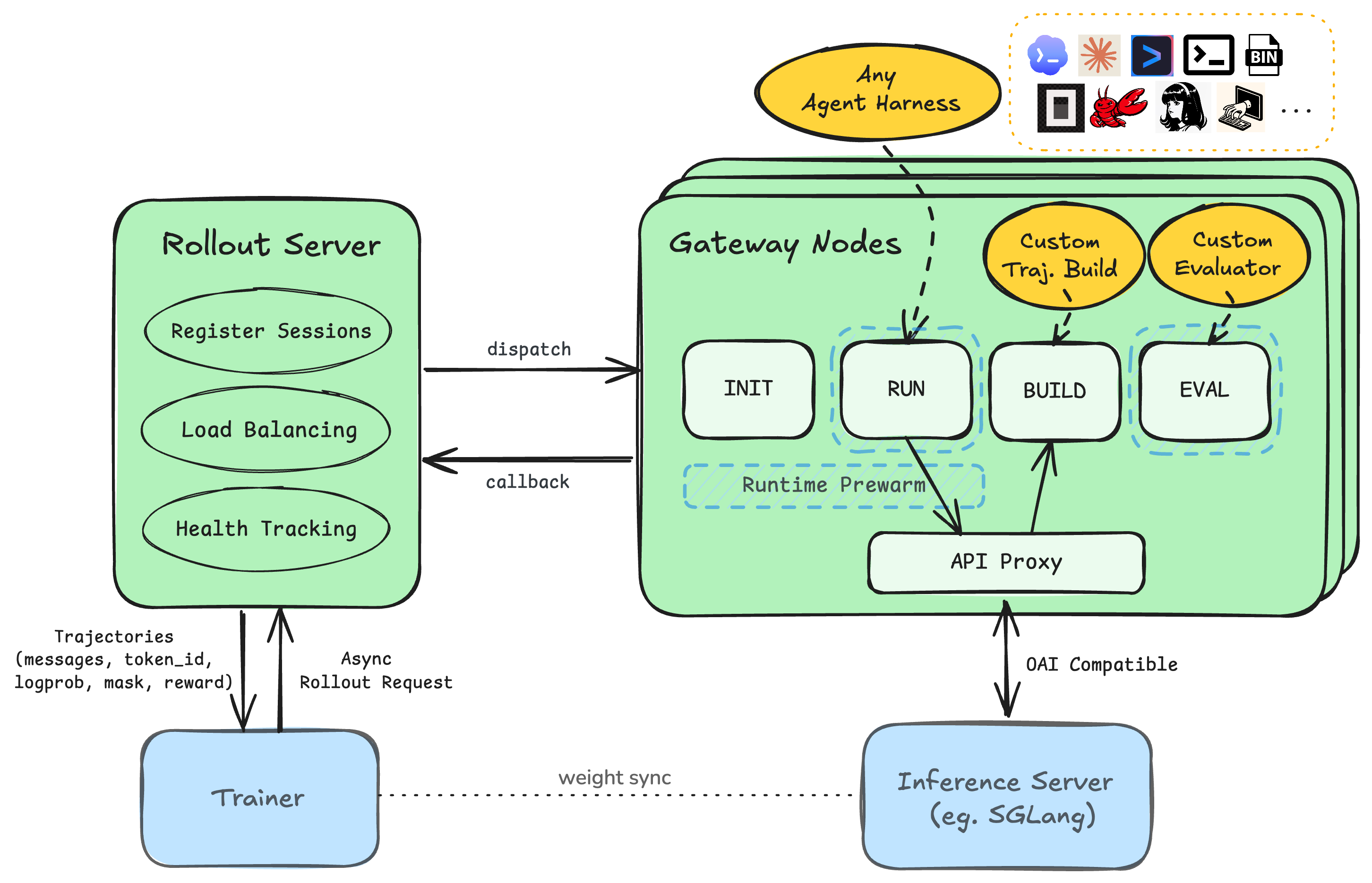}
\caption{\textbf{\polar architecture overview.} \polar runs an existing agent harness inside an isolated runtime and places a model API proxy between the harness and the inference server. The proxy forwards model calls, records token-level request and response data, and reconstructs RL trajectories, while rollout gateways asynchronously handle runtime prewarming, harness execution, evaluation, and trainer callbacks. This decoupled design allows \polar to treat agents as a black-box environment, seamlessly scaling across different training frameworks.}
\label{fig:polar_architecture_overview}
\label{fig:polar_architecture}
\end{figure}

\section{Introduction}

Reinforcement learning for large language model is moving beyond short, single-step tasks toward agentic settings~\citep{prorlagent2026,rllm2025} that require sustained interaction with external environments, such as code repositories~\citep{swebench2024,swegym2024}, web browsers~\citep{zhou2023webarena,deng2023mind2web}, and even full operating systems~\citep{xie2024osworld,wang2025opencua}, through iterative tool use~\citep{shao2024deepseekmath,guo2025deepseek,patil2025the}.
These settings often produce long-horizon trajectories with dozens of interaction steps and tens of thousands of tokens.

\begin{figure}[t]
\centering
\includegraphics[width=\textwidth]{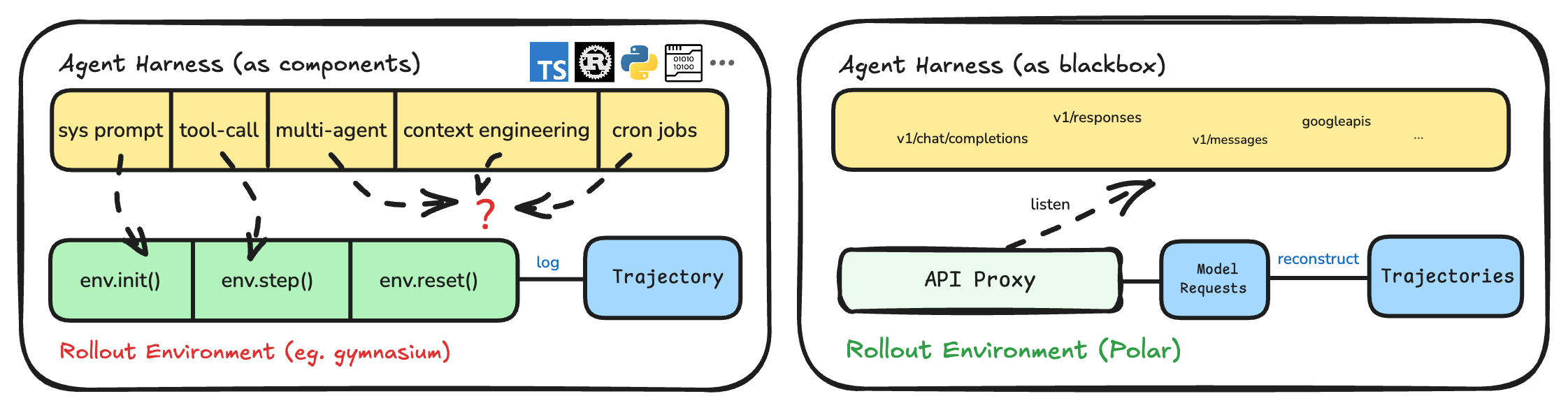}
\caption{\textbf{\polar uses the model API proxy as the rollout boundary.} Traditional rollout frameworks usually require the agent or harness logic to be rewritten behind a framework-owned environment API. This makes the trainer depend on harness-specific integration code and can miss details of the native execution path. \polar instead keeps the harness unchanged and places a provider-compatible proxy at the LLM API boundary; the proxy records prompts, sampled tokens, log probabilities, and responses, then reconstructs trainer-ready trajectories outside the harness.}
\label{fig:rollout_style_comparison}
\end{figure}

This shift makes the training target itself a central systems challenge for agentic RL. Traditional RL often assumes the training target can be exposed through a simple, standardized interface~\citep{brockman2016openai}, allowing researchers to focus mainly on the RL algorithm.
In agentic RL, however, the training target is often a complex software system~\citep{openai_codex_2026, anthropic_claude_code_2026}. It may involve heterogeneous environments~\citep{nemo-gym}, various external tools~\citep{zhang2026nemotronresearchtooln,zhang2024offline}, and long-running workflows~\citep{swebench2024}, and may be implemented in different languages or even distributed as a closed-source binary~\citep{anthropic_claude_code_2026}.

This creates substantial integration burdens. For example, in building agentic RL systems, SkyRL-Agent~\citep{skyrlagent2025} and PRIME-RL~\citep{primerl2026} integrate agent execution directly into the RL pipeline, requiring users to adapt their agents to the RL infrastructure rather than allowing the infrastructure to accommodate existing agent implementations.
This makes the design less flexible: every new agent or harness often requires one framework-specific integration.
Some recent systems attempt to make this integration less intrusive. For instance, Agent Lightning~\citep{agentlightning2025} and rLLM~\citep{rllm2025} reduce this burden by introducing standard tracing interfaces and LLM-call capture mechanisms, but still require agents to conform to prescribed interfaces. Thus, these systems lower the cost of integration but do not fully eliminate it.
These issues are likely to become more severe as agent harnesses grow increasingly complex and, in some cases, even do not expose their internal implementation, making conventional RL integration difficult or even infeasible. Motivated by
these issues, we explore the following central question:

\begin{center}
    \emph{Can we train agents with RL without opening the box?}
\end{center}

That is, without touching their harnesses or forcing them to conform to an RL framework.
The key observation is that, although agents differ widely in their internal implementations, every LLM-based agent must talk to a model. This model API boundary provides a common interface that exists outside the agent itself.
Instead of integrating with the agent harness, we can train by listening to the agent’s LLM calls: capturing its prompts, sampled tokens, log probabilities, and responses, and converting them into RL trajectories. In this view, an agent can be treated as a black box while still becoming trainable.

Building on this intuition, we present \polar: an agentic RL infrastructure that could train \textbf{any} agents as black boxes. The name \polar reflects both its roots in \textbf{P}r\textbf{O}r\textbf{L} \textbf{A}gent serv\textbf{R}~\citep{prorlagent2026} and its role in connecting the two ``poles'' of agent training and deployment: the training environment and the product harness. Instead of treating the agent harness as the RL interface, \polar uses the agent’s LLM API traffic as the interface. Through listening to its model calls through a proxy and converts them into trajectories and rewards for training, the agent runs unchanged.

In addition, \polar separates runtime setup, agent execution, trajectory reconstruction, evaluation, and trainer callbacks behind asynchronous service boundaries. This allows slow and long-tail agent rollouts to scale independently from GPU training, exposing a trainer-agnostic rollout-as-a-service interface for scaling efficient RL infrastructures~\citep{prorlagent2026}.
In summary, the main contributions of this work are:
\begin{itemize}
    \item \textbf{Proxy based rollout and reconstruction over agent harnesses.}
    We propose a paradigm using the agent's LLM API payloads as the RL rollout interface, allowing existing harnesses to serve directly as RL environments without internal code change.
    
    \item \textbf{Rollout-as-a-service architecture for scaling RL infrastructures.}
    \polar separates task submission, runtime setup, harness execution, trajectory reconstruction, evaluation, and trainer callbacks behind asynchronous service boundaries, natively scaling with modern RL infrastructures.

    \item \textbf{Token-faithful trajectory reconstruction.}
    \polar converts raw model requests into token-faithful traces for training. We provide conservative per-request reconstruction and prefix merging for heavy rolllouts, while leaving registry-based extensible interfaces.

    \item \textbf{End-to-end validation on real-world coding harnesses.}
    We validate \polar with RL training on various popular harnesses for software-engineering tasks, and further demonstrate offline SFT data generation with a custom coding harness.
\end{itemize}

\section{Related Work}

A compact checklist of rollout-system design choices is provided in \cref{tab:framework_comparison} in the appendix. This section focuses on the qualitative differences behind that comparison.

\subsection{Agent RL Systems}

The first wave of LLM RL infrastructure largely assumed that rollout generation was a Python function owned by the trainer. This assumption is increasingly strained by multi-turn agents, where interaction spans many model calls and environment actions. \prorlagent \citep{prorlagent2026} introduced a service boundary for multi-turn agent rollouts, separating sandbox setup, agent execution, and reward computation from the training process. \polar inherits the same high-level idea that rollout should be a service, but changes the integration contract. Instead of implementing an agent handler inside the rollout service, the user supplies a harness adapter that prepares configuration and launches the native executable. The model proxy then observes the harness from outside.

SkyRL-Agent \citep{skyrlagent2025} is a full-stack system for efficient RL training and evaluation of multi-turn, long-horizon agents, with SkyRL-Gym providing tool-use environments through a Gymnasium-style interface. SkyRL's strength is efficient training once tasks are represented in its environment and agent abstractions. \polar is complementary: it targets the earlier systems problem of running a pre-existing harness whose internal event loop, tool formatting, and context policy should remain unchanged.

PRIME-RL \citep{primerl2026} focuses on large-scale asynchronous RL with trainer-inference separation, stale-policy step semantics, and support for verifiers environments. Slime \citep{slime2025,sglang2024} similarly connects Megatron training with SGLang rollout engines and exposes customizable data-generation interfaces. These systems address the policy-optimization and inference-scaling side of the pipeline. \polar is not a replacement trainer. It is a rollout substrate that can feed asynchronous trainers with trajectories from heavier harnesses than typical verifiable-reward functions.

\subsection{Low-Intrusion Agent Instrumentation}

Agent Lightning \citep{agentlightning2025} proposes a training-agent disaggregation architecture and a unified data interface for converting agent execution into trainable transitions. rLLM \citep{rllm2025} similarly aims to train agents across frameworks with minimal code changes, using tracked clients, decorators, workflow abstractions, and proxy support to collect token IDs and log probabilities. Both systems recognize that researchers should not have to rewrite complete applications to train them.

\polar differs in the chosen minimum integration point. For many coding and terminal agents, the most reliable interface is not an SDK callback graph but the provider API endpoint already used by the harness. The gateway proxy therefore becomes the observation device: it accepts Anthropic, OpenAI Chat, OpenAI Responses, and Google-style requests; translates them to the local inference backend; and records the token-level fields needed by the trainer. This choice is narrower than general observability instrumentation, but it is robust to harnesses implemented as command-line programs, package-managed tools, or binaries.


\subsection{SWE Task Evaluation and Benchmark}

Harbor \citep{harbor2026} evaluates agents such as Claude Code, OpenHands, Codex CLI, and related systems in containerized environments, supports parallel execution through local and cloud providers, and converts native agent logs into evaluation trajectories. This evaluation-first design is highly aligned with \polar's harness-native motivation. The difference is the model boundary and training data contract. Harbor launches each harness with provider-specific configuration and does not provide a gateway that translates model-provider protocols or mediates the harness's model traffic. As a result, model substitution is limited by what the native harness and external endpoint already support: for example, evaluating a Qwen checkpoint through Claude Code requires an Anthropic-compatible endpoint outside Harbor. \polar instead places a proxy at this boundary, so the same style of harness execution can yield token IDs, log probabilities, loss masks, and rewards that are directly consumable by an RL trainer.

SWE-bench \citep{swebench2024} established real GitHub issue resolution as a benchmark requiring repository understanding, editing, and executable validation. SWE-Gym \citep{swegym2024} extends this direction with training environments, verifiers, and trajectories for software-engineering agents. These workloads are a natural stress test for rollout infrastructure because they combine expensive runtime setup, sparse patch-level rewards, long-tail execution time, and many opportunities for harness-side state to diverge from a clean evaluator state.

\subsection{Token Fidelity and Retokenization Drift}

The training signal in agent RL is only correct if it is attached to the tokens sampled by the behavior policy. This is difficult in agent harnesses because provider APIs may return text, tool-call JSON, reasoning fields, or streamed events rather than the exact token IDs and log probabilities used by the inference backend. The vLLM and Agent Lightning discussion of retokenization drift emphasizes that decoding and re-encoding a transcript can produce different token IDs from the original generation \citep{vllmretokenization2025}. \polar follows the same token-fidelity principle but applies it to arbitrary harness rollouts: generated assistant tokens are copied from inference responses, non-generated interstitial tokens are taken from canonical prompt tokenization, and the loss mask marks only behavior-policy tokens as trainable.

\section{Polar}
\label{sec:method}

We target agentic RL tasks where a policy is exercised through an existing harness rather than a custom rollout loop. A task starts from an instruction and a runtime; the harness calls a model endpoint while using tools, editing files, spawning sub-agents, or managing context (compaction, injection, replacement, etc.). After execution, an evaluator assigns an outcome or trace-level reward. The rollout system must preserve the native model interactions as trainer-ready traces: prompt context, sampled assistant tokens, optional behavior-policy log probabilities, loss masks, rewards, and provenance.

\subsection{Architecture}

\polar has two core components: a rollout server and gateway nodes. The rollout service coordinates tasks and global scheduling. Gateway nodes execute sessions, host the model proxy, construct trajectories, and run evaluation. This split keeps durable task management separate from per-session execution and capture.

\begin{figure*}[t]
\centering
\includegraphics[width=0.86\textwidth]{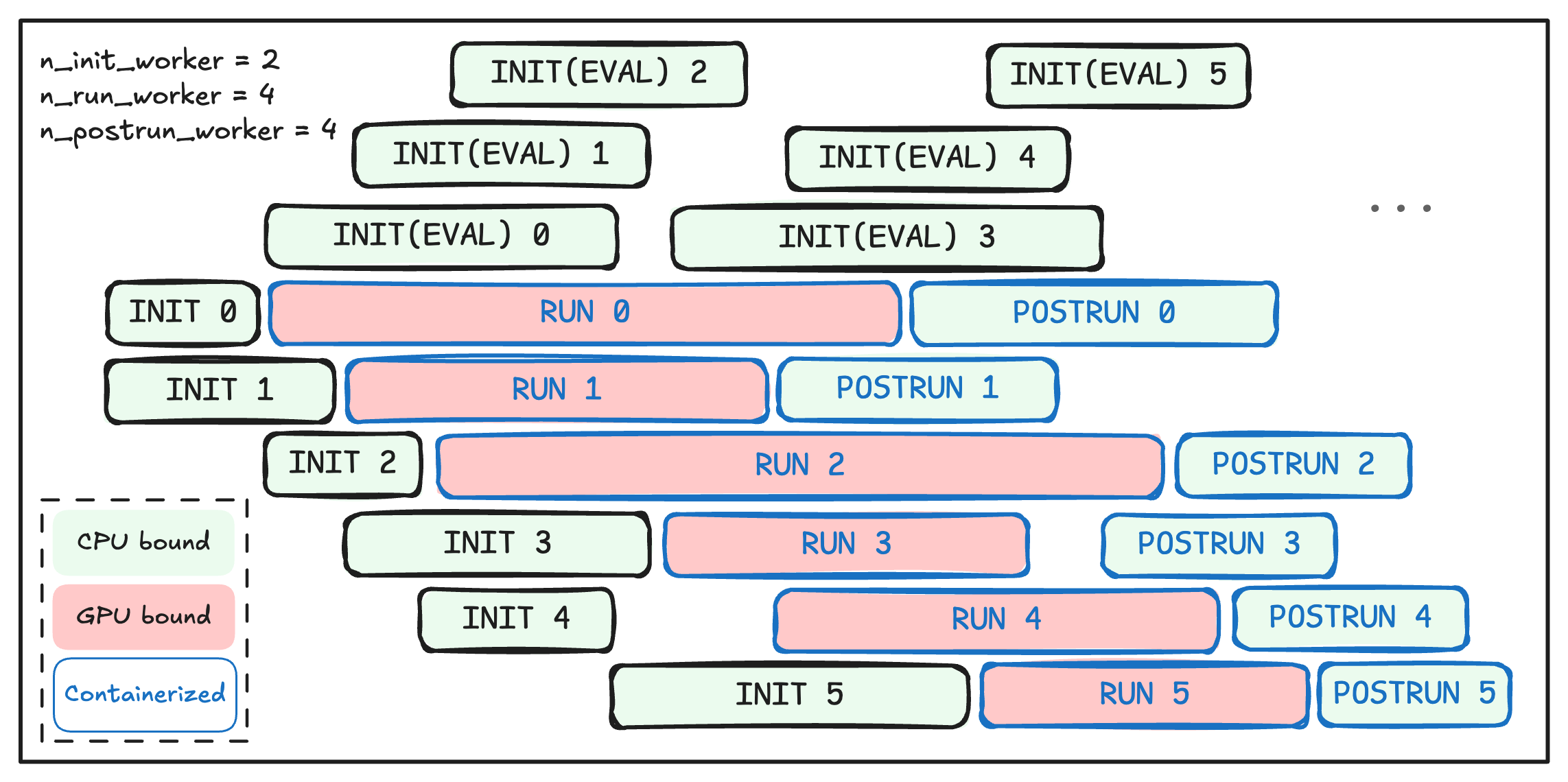}
\caption{\textbf{Gateway-level asynchronous staging in \polar.} A gateway separates runtime initialization, ready buffering, harness execution, and post-run trajectory and evaluation work into isolated worker pools. Runtime preparation and evaluator prewarm proceed off the critical path, so CPU-heavy runtime setup and long-tail evaluation do not block active GPU-bound agent run.}
\label{fig:inner_async_staging}
\label{fig:async_staging}
\end{figure*}

\paragraph{Rollout server.}
The rollout service accepts a \texttt{TaskRequest} and expands it into \texttt{num\_samples} independent sessions. A session is the scheduling unit: it has a session ID, task ID, timeout budget, runtime specification, agent specification, trajectory builder, evaluator, and callback URL. The service dispatches sessions to gateway nodes, persists compact terminal results, exposes task status through polling, and accepts gateway callbacks when sessions finish. \Cref{app:task_payload} gives a representative payload.

\paragraph{Gateway node.}
A gateway owns the lifecycle of each session. It starts the runtime, prepares the harness, runs the harness commands, builds trajectories from captured completions, evaluates the output, tears down resources, and returns the result. The same gateway also hosts the proxy endpoint used by the harness for model calls. This co-location keeps completion capture tied to the session registry and avoids a separate trace-collection service.

Training frameworks are independent from \polar servers. And the service boundaries natively supports efficient asynchronous RL at scale. \Cref{fig:outer_async_rl} shows one such example with Slime: a background worker submits \polar tasks, receives task-completion callbacks, converts traces into Slime \texttt{Sample} objects, and applies trajectory-aware reward post-processing.

\subsection{Harness and Proxy Capture}

\polar observes native harnesses by routing their model calls through the gateway proxy. A harness is configured through its normal environment variables or config files so that its model base URL points to the gateway.

For each incoming model request, the gateway performs four steps.

\begin{enumerate}
    \item \textbf{Detect the provider API.} Detection uses the request path and headers to distinguish Anthropic Messages, OpenAI Chat Completions, OpenAI Responses, and Google \texttt{generateContent}-style calls.
    \item \textbf{Normalize the request.} A provider transformer converts roles, content parts, tool definitions, tool choices, stop controls, and generation parameters into the OpenAI Chat Completions shape consumed by local inference servers. The transformer also adds fields needed for training, such as \texttt{logprobs=true}.
    \item \textbf{Capture token-level data.} The gateway forwards the normalized request to the inference servers and stores a completion record containing the request messages, response messages, prompt token IDs, sampled response token IDs, finish reason, and log probabilities from inference backends.
    \item \textbf{Return the provider shape.} The response is transformed back to the schema expected by the harness. For streaming requests, our implementation obtains a non-streaming upstream response and emits a synthetic provider-shaped stream. This simplifies faithful token capture while preserving compatibility with harnesses that expect server-sent events.
\end{enumerate}

The proxy boundary is intentionally below the agent framework. It does not need to understand how the harness plans, manages tools, or decides when to stop. It only needs to preserve API compatibility and record enough information to reconstruct training samples.

\subsubsection{Harness Adapter}

A harness adapter in \polar is small by design. It may install configuration, register MCP servers or skills, write provider settings, and return the shell commands that run the agent. 
A generic \texttt{shell} command harness can be used for wrapped agent execution. We also integrate popular agent harnesses as shortcuts like \texttt{claude\_code}, \texttt{codex}, \texttt{gemini\_cli}, \texttt{qwen\_code}, \texttt{opencode} and \texttt{pi}.

\subsubsection{Runtime Interface}

Runtimes implement a common interface for \texttt{start}, \texttt{stop}, \texttt{exec}, upload, download, and cancellation. Our first release supports Docker and rootless Apptainer for HPC setup. Because gateway code only depends on the runtime interface, a task can change isolation backend without friction.

\subsection{Asynchronous Rollout Staging}

Long-horizon harness rollouts mix several different costs: runtime startup, dependency preparation, harness execution, evaluator setup, test execution, patch application, and teardown. \polar keeps these costs from blocking one another through stage-isolated execution inside each gateway (\cref{fig:async_staging}).

\subsubsection{In-node worker pools.}
Each gateway uses isolated worker pools for \texttt{INIT}, \texttt{RUNNING}, and \texttt{POSTRUN}, plus a bounded \texttt{READY} buffer. \texttt{INIT} starts the runtime and executes prepare actions. \texttt{READY} holds initialized runtimes until a run slot is available. \texttt{RUNNING} executes the harness. \texttt{POSTRUN} builds trajectories, runs evaluators, executes post-run hooks, sends callbacks, and tears down resources. The ready buffer allows CPU-heavy runtime preparation to proceed in the background without blocking GPU-bound agent execution.


\subsubsection{Evaluator prewarm and timeouts.}
When an evaluator requests a clean runtime, the gateway begins preparing that runtime during the agent run. Each session also carries one shared deadline; if a harness times out after model calls have been captured, the gateway still enters post-run so partial traces can be recovered with terminal timeout status.

\begin{figure*}[t]
\centering
\includegraphics[width=0.76\textwidth]{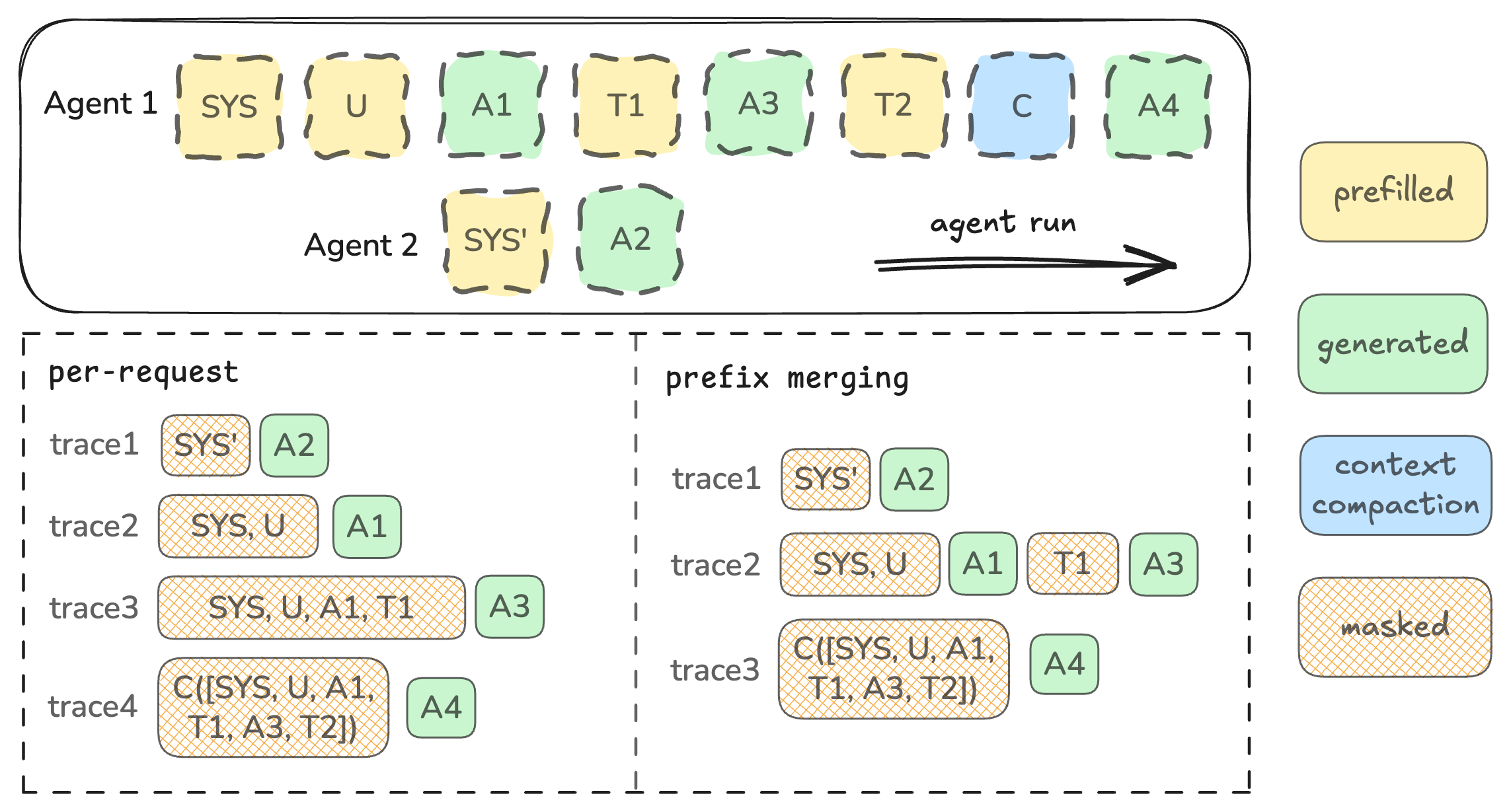}
\caption{\textbf{Trajectory reconstruction example.} The visualized session contains a three-turn main agent that undergoes one harness-level context compaction and spawns one subagent. The per-request builder keeps each captured model call as an independent trace. Prefix merging instead recovers append-only conversation chains where valid, while compaction and subagent boundaries naturally form separate chains. Within each merged trace, \polar prefix merging algorithm copies only sampled assistant tokens as trainable tokens and masks canonical interstitial tokens, preserving behavior-policy fidelity while reducing trainer-facing samples.}
\label{fig:trajectory_builder_outputs}
\label{fig:trajectory_builders}
\end{figure*}

\begin{figure*}[t]
\centering
\begin{subfigure}[c]{0.48\textwidth}
\centering
\includegraphics[width=\linewidth]{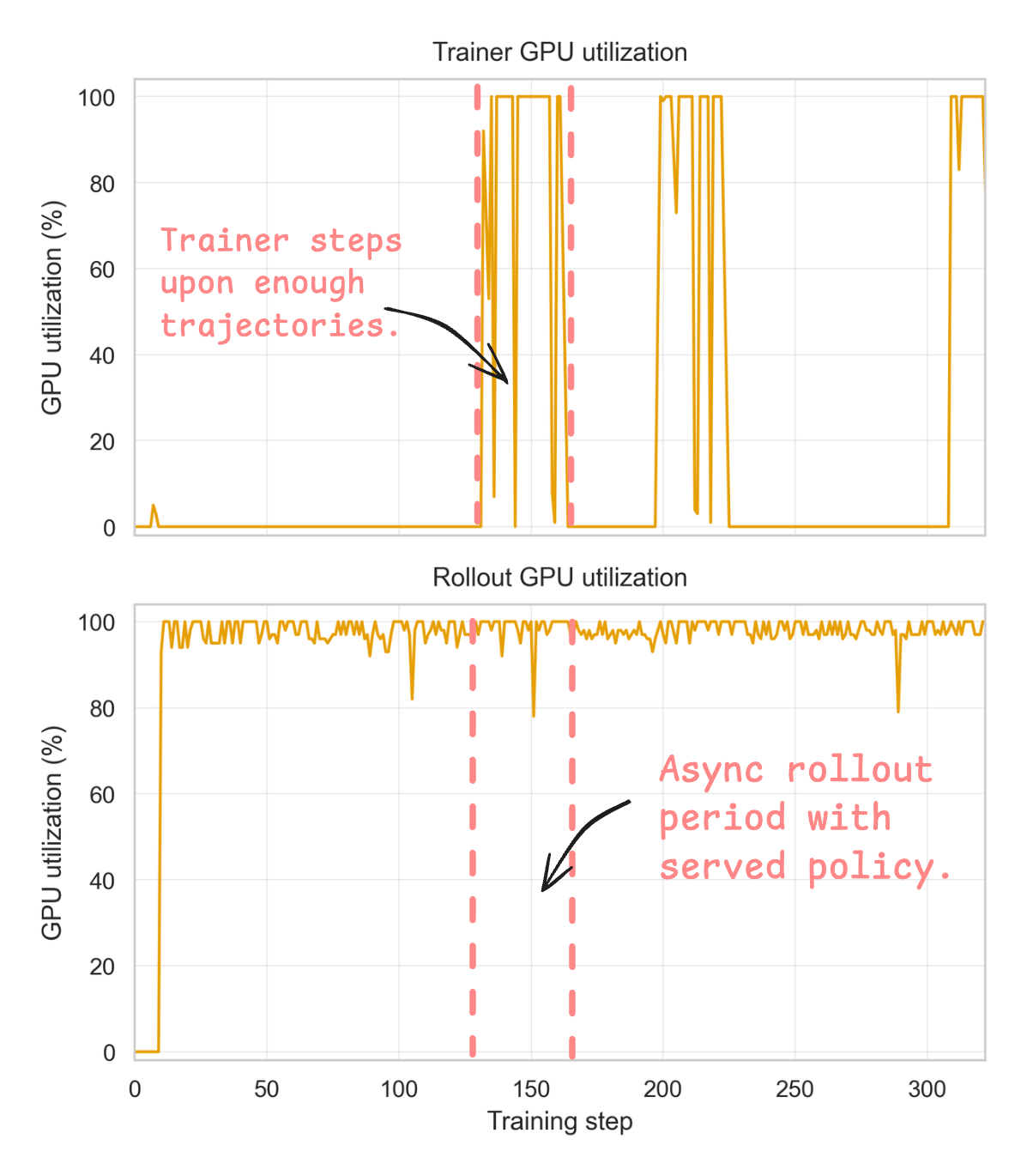}
\caption{Async RL.}
\label{fig:outer_async_rl}
\end{subfigure}
\hfill
\begin{subfigure}[c]{0.48\textwidth}
\centering
\includegraphics[width=\linewidth]{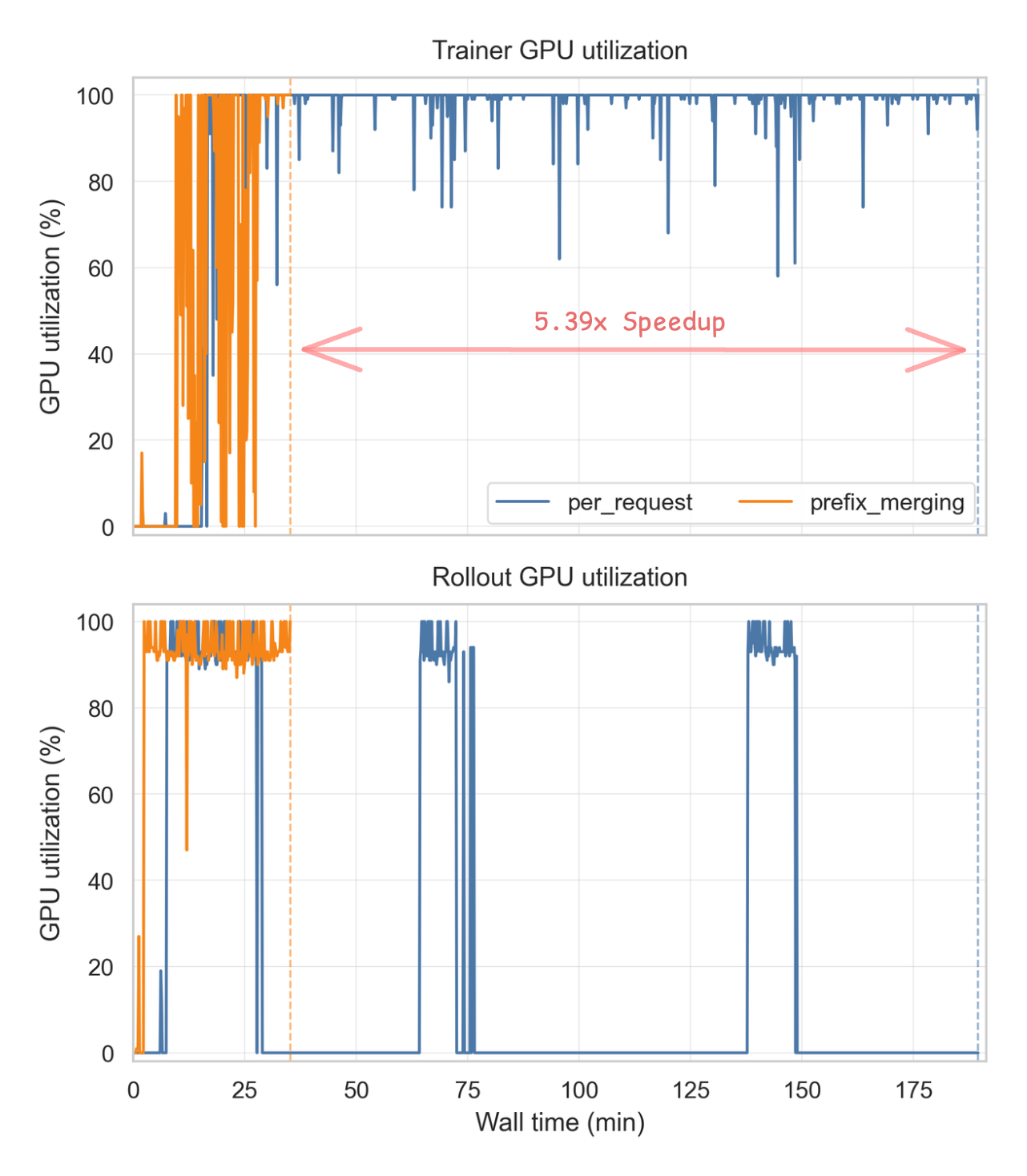}
\caption{GPU utilization with different reconstruction strategies.}
\label{fig:builder_gpu_utilization}
\end{subfigure}
\caption{\textbf{\polar improves GPU utilization across the rollout-training boundary.}
\captiona{} Shows an asynchronous RL pipeline enabled by \polar services. The rollout server keeps inferencing with existing policy, while trainer steps only if receiving batch size of evaluated trajectory groups.
\captionb{} Shows a span of 3 training steps under the same workload and topology, prefix merging emits fewer trainer updates than per-request reconstruction and substantially accelerates the training process.}
\label{fig:rollout_training_utilization}
\end{figure*}

\subsection{Trajectory Reconstruction}

The trajectory builder interface converts an ordered \texttt{CompletionSession} into a \texttt{Trajectory}. A completion session is the stored sequence of proxy-captured model calls for one harness session. A trajectory contains one or more \texttt{Trace} objects, each with prompt token IDs, response token IDs, a loss mask, prompt messages, response messages, tool definitions, log probabilities, reward, and metadata. \Cref{app:trace_example} shows a representative trainer-facing trace.
Custom trajectory strategies can be seamlessly added to the registry, and we provide two strategies, per request and prefix merging, compared in \cref{fig:trajectory_builders}.

\subsubsection{Per Request}

The \texttt{per\_request} builder is the conservative baseline as shown in the bottom left of  \cref{fig:trajectory_builders}: every completion becomes one trace. This is lossless with respect to individual calls, but it can fragment a coherent multi-turn agent session into many short samples. For complex coding harnesses, solving a single coding problem can produce hundreds of such traces, which increases the burden on downstream trainers.

\subsubsection{Token Faithful Prefix Merging}
\label{sec:prefix_merging}

The \texttt{prefix\_merging} builder reconstructs longer traces when parts of a harness session preserve append-only conversation histories as shown in the bottom right of \cref{fig:trajectory_builders}. It does not assume that the whole session is a single conversation. Instead, for a session with completions $C_1,\ldots,C_T$, where completion $C_i$ has prompt token sequence $p_i$, raw sampled response token sequence $a_i$, response log probabilities $\ell_i$, and prompt/response messages $m_i$, \polar partitions the completions into ordered chains
\[
\mathcal{G} = \{G_1,\ldots,G_J\}, \qquad
G_j = (C_{i^j_1}, C_{i^j_2}, \ldots, C_{i^j_{K_j}}),
\]
with $i^j_1 < i^j_2 < \cdots < i^j_{K_j}$. A new completion can join an existing chain only when a normalized message-level grouping key identifies it as a candidate continuation and the strict token-prefix relation holds against the last prompt in that chain. For adjacent completions $C_{i_m}$ and $C_{i_{m+1}}$ inside one chain, this check is
\[
p_{i_{m+1}}[1:|p_{i_m}|] = p_{i_m}.
\]
Thus sub-agents, parallel agent branches, context compaction, prompt rewriting, or independent tool-mediated conversations naturally form additional chains rather than being forced into one global trace.

Merging is then applied independently to each chain. Consider one chain $G=(C_{i_1},\ldots,C_{i_K})$ and write $p_m=p_{i_m}$, $a_m=a_{i_m}$, and $\ell_m=\ell_{i_m}$. The main challenge is that $p_{m+1}$ contains a canonical server rendering of the previous assistant turn plus the interstitial context inserted by the harness before the next generation prompt. The previous assistant body must not be copied from this canonical rendering, because the behavior-policy tokens are the raw sampled tokens $a_m$. Let $e$ denote the end-of-turn token ID. For two adjacent completions in the chain, define the canonical tail
\[
t_m = p_{m+1}[|p_m|+1:].
\]
We locates the first $e$ in $t_m$. If $a_m$ already ends with $e$, the interstitial $u_m$ is the suffix after that $e$; otherwise $u_m$ starts at that $e$ so the assistant turn is still closed before the next prompt context. The token sequence represented by this chain is
\[
z^{(j)} = p_1 \; || \; a_1 \; || \; u_1 \; || \; a_2 \; || \; u_2 \; || \cdots || \; a_K .
\]
The emitted trajectory therefore contains one trace $\tau^{(j)}$ per chain, with the first prompt $p_1$ stored as the trace prompt and the remaining suffix $a_1 || u_1 || \cdots || a_K$ stored as the trace response. The explicit loss mask is one on tokens copied from sampled responses $a_m$ and zero on tokens copied from canonical interstitials $u_m$. Real response log-probability entries are copied for $a_m$ tokens. Interstitial slots receive synthetic log-probability entries so \texttt{response\_logprobs} stays aligned with \texttt{response\_ids}; trainability is controlled by \texttt{loss\_mask}.

This construction gives a simple correctness invariant in every emitted trace:
\begin{center}
    \emph{Every trainable token matches the behavior policy during rollout, and any non-generated tokens are masked out.}
\end{center}


\cref{fig:builder_gpu_utilization} compares the GPU utilizations of the 2 strategies above with the same configurations.

\subsection{Evaluation and Reward Propagation}

Evaluators are registry-backed custom strategies that run after trajectory construction. They receive the trajectory, session artifacts, and optionally refreshed runtime context. Built-in evaluators include a session-completion reward, a configurable test-on-output evaluator, and a SWE-Bench/SWE-Gym harness evaluator. An outcome reward can be broadcast to every trace, whereas tasks with process rewards may need per-trace assignment.
The evaluator registry allows straightforward extension to custom rule-based verification, agent-as-judge scoring, and task-specific reward shaping.

\section{Experiments}
\label{sec:experiments}

We validate \polar in two settings: online RL rollout and offline SFT data generation.
The experiments test whether unchanged harnesses can produce trainable traces for both reward-driven and supervised training.

\subsection{SWE-Gym GRPO on Coding Harnesses}

\begin{figure}[t]
\centering
\includegraphics[width=\textwidth]{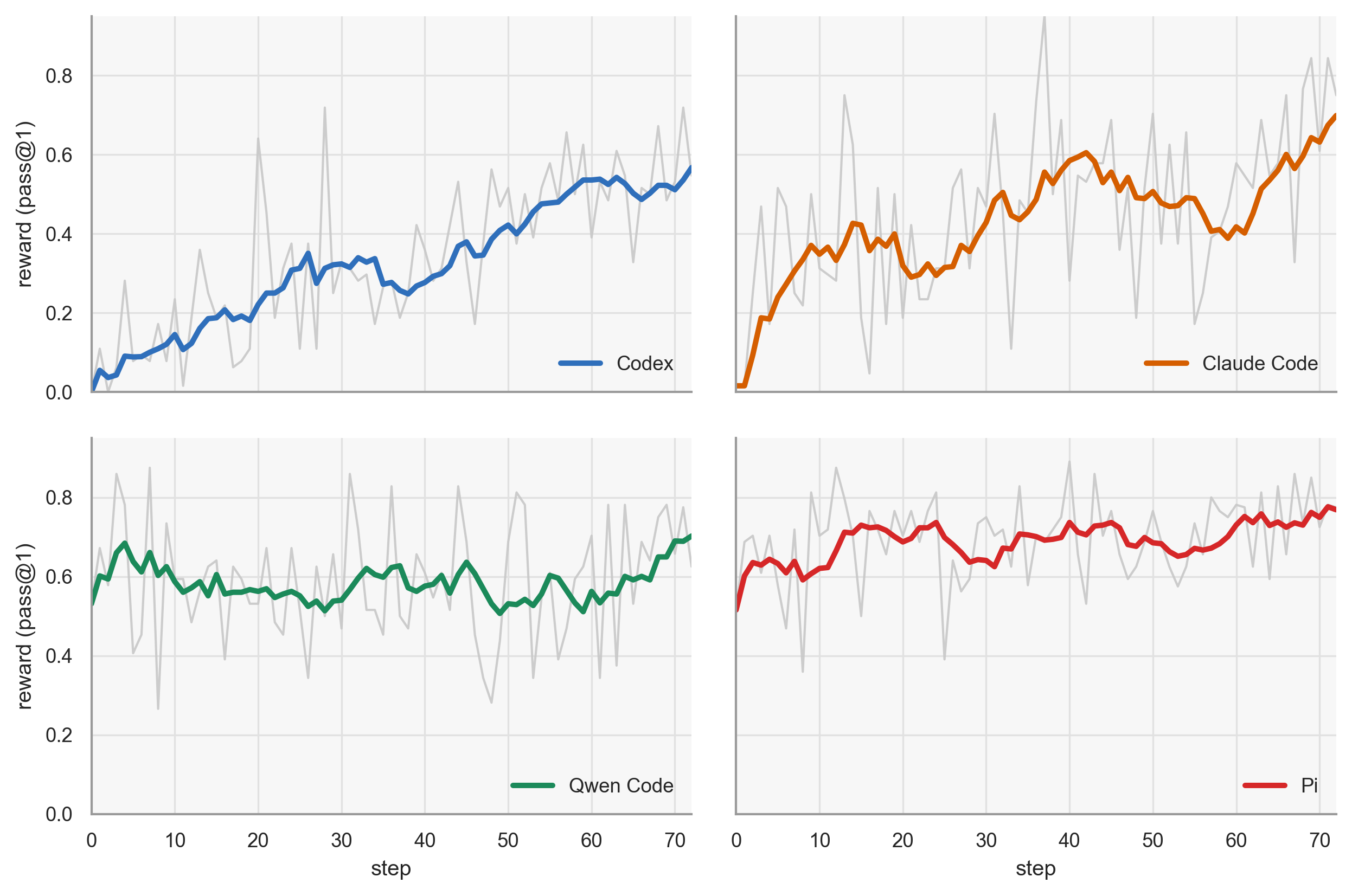}
\caption{\textbf{SWE-Gym GRPO training curves.} Each panel shows the per-step outcome reward, equivalent to rollout pass@1, for one of four evaluated coding harnesses. RL improves reward across harnesses, with the clear gains on execution paths involving complex prompting, orchestration, or unfamiliar tool schemas.}
\label{fig:swegym_grpo_training_curves}
\end{figure}

We run standard GRPO over four representative coding harnesses with \polar. Starting from the same Qwen3.5-4B base checkpoint, we run standard GRPO training on the \texttt{SkyRL-v0-293-data} SWE-Gym dataset,\footnote{\url{https://huggingface.co/datasets/NovaSky-AI/SkyRL-v0-293-data}} with \polar and Slime.
We use the training split for policy optimization and reserve evaluation for SWE-Bench Verified~\citep{swebench2024}.
All experiments use \texttt{prefix\_merging} to convert llm traffics from harnesses into trainable traces. And we use \texttt{swebench\_harness} to score the final edition patch in a fresh runtime. The main training hyperparameters are listed in \cref{tab:swegym_grpo_hparams}.

\cref{fig:swegym_grpo_training_curves} shows the training reward for Codex, Claude Code, Qwen Code, and Pi. The Codex run begins near zero reward and rises steadily over training, with the last ten steps averaging 54.5\% pass@1 reward compared with 9.5\% over the first ten steps. Claude Code also improves substantially, rising from 28.8\% over the first ten steps to 67.0\% over the last ten steps. Qwen Code and Pi start from stronger native-harness priors: Qwen Code is noisier but rises from 61.6\% to 66.0\%, while Pi improves more clearly from 61.6\% to 76.2\% over the same first and last ten-step windows. These curves are consistent with the benchmark result in \cref{tab:swebench_verified_eval}: \polar improves the same 4B base model under all four evaluated harnesses. The largest absolute gain appear in Codex, likely due to unfamiliar tool schemas.

\begin{table}[t]
\centering
\small
\setlength{\tabcolsep}{8pt}
\begin{tabular}{lccc}
\toprule
\textbf{Harness} & \textbf{Base} & \textbf{\polar RL} & \textbf{Gain} \\
\midrule
Codex & 3.8\% & \textbf{26.4\%} & \textbf{22.6 $\uparrow$} \\
Claude Code & 29.8\% & \textbf{34.6\%} & \textbf{4.8 $\uparrow$} \\
Qwen Code & 34.6\% & \textbf{35.2\%} & \textbf{0.6 $\uparrow$} \\
Pi & 34.2\% & \textbf{40.4\%} & \textbf{6.2 $\uparrow$} \\
\bottomrule
\end{tabular}
\caption{\textbf{SWE-Bench Verified evaluation.} All rows start from the same Qwen3.5-4B base model and are trained over listed harnesses. Scores are pass@1 over the full benchmark, running on corresponding harnesses.}
\label{tab:swebench_verified_eval}
\end{table}

The evaluation isolates the value of harness-native RL. Under Codex, the 4B base model reaches only 3.8\% pass@1 before training, but the \polar-trained checkpoint reaches 26.4\%, a 22.6 point absolute gain. This large jump is expected: Codex presents an unfamiliar action protocol, context policy, and patch-submission style to a Qwen model that was not originally trained as a Codex-native policy. \polar keeps that harness unchanged and attaches the reward to the actual sampled tokens flowing through the Codex execution path, so GRPO optimizes the behavior the model must use at evaluation time. Under the native Qwen Code harness, the base model is already much stronger at 34.6\%, and \polar still improves it to 35.2\%, a 0.6 point gain. Claude Code also improves from 29.8\% to 34.6\%, adding 4.8 points, and Pi improves from 34.2\% to 40.4\%, adding 6.2 points. These results show that harness-native RL can deliver large adaptation gains for unfamiliar execution paths while still preserving gains when the base checkpoint is already well aligned with the harness.

\paragraph{Trajectory builder ablation.}

We ablate the trajectory builder under identical model, hardware, and topology settings, changing only whether captured completions are emitted as \texttt{per\_request} traces or merged by \texttt{prefix\_merging}.
\cref{fig:builder_gpu_utilization} shows a partial utilization profile.
Over same three training steps, \texttt{prefix\_merging} reduces the trainer stream from 1{,}185 request-level updates to 218 merged-trace updates, cutting wall-clock time from 189.5 to 35.2 minutes (5.39$\times$).
\texttt{prefix\_merging} keeps rollout GPUs active with 87.7\% average rollout utilization, compared with 20.4\% average utilization for \texttt{per\_request} over the same period.

We also tried \texttt{per\_request} with outcome-reward broadcasting to every emitted trace, but observed significant reward hacking. The issue is noisy credit assignment: request-level traces can receive session-level credit without proper session normalization or an advanced process reward model. Those mechanisms are outside the scope of this work, but providing examples and tools for session normalization and PRM-style credit assignment is on our roadmap.

\subsection{Offline Data Generation}
\label{sec:offline-data-gen}

Beyond serving online RL rollouts, \polar can be repurposed as a distributed
\emph{offline} data-generation service: a fixed checkpoint and harness are
fanned out across the cluster, every session is journaled to disk, and the
resulting traces are filtered and post-processed for downstream training.
The same primitives that make \polar useful for RL---per-session container
isolation, automatic retry, and gateway-mediated scheduling.

\paragraph{Case study: SWE-Gym SFT trajectories.}
We used \polar to generate a supervised fine-tuning corpus of agentic
software-engineering trajectories. The setup is intentionally minimal: a single
8$\times$H100 SGLang serve job hosting Qwen3.5-122B-A10B (TP$=$8,
\texttt{max\_model\_len}$=$32{,}768) drives the
\texttt{pi-coding-agent}~v0.67.68 harness against 1{,}638 instances drawn from
seven SWE-Gym repositories. Each task runs in its own Apptainer SIF built from
the SWE-Gym reference image with Node.js~22 and the harness layered on top, so
the agent's tool calls (\texttt{bash}, \texttt{read}, \texttt{edit},
\texttt{write}) execute against a fresh checkout of the target commit. Submission
uses \texttt{max\_concurrent}=5--8, \texttt{max\_retries}=1, and a per-task
timeout of 3{,}600 seconds; trajectories that finished with
\texttt{empty\_generation} are retried once and the rest accepted as-is.

A trajectory is accepted into the SFT corpus if and only if the SWE-Bench
evaluation harness reports the agent's final patch as resolving every
\texttt{FAIL\_TO\_PASS} test while leaving every \texttt{PASS\_TO\_PASS} test
green. With this single-bit filter, \polar produced \textbf{504 accepted
trajectories from 1{,}638 attempts} (30.8\% acceptance), at a cost of roughly
64 GPU-hours on the interactive partition. Per-repository acceptance varies
substantially with task difficulty (Table~\ref{tab:swegym-pi-rates}): bug-fix
heavy repositories like \texttt{getmoto/moto} accept at over 50\%, while
data-frame and dataflow workloads with longer test suites accept below 20\%.

\begin{table}[t]
\centering\small
\begin{tabular}{lrrr}
\toprule
\textbf{Repo} & \textbf{Attempts} & \textbf{Accepted} & \textbf{Rate} \\
\midrule
\texttt{getmoto/moto}      & 343 & 184 & 53.6\% \\
\texttt{python/mypy}       & 257 & 101 & 39.3\% \\
\texttt{conan-io/conan}    &  71 &  27 & 38.0\% \\
\texttt{pydantic/pydantic} &  81 &  24 & 29.6\% \\
\texttt{iterative/dvc}     & 219 &  45 & 20.5\% \\
\texttt{pandas-dev/pandas} & 477 &  98 & 19.7\% \\
\texttt{dask/dask}         & 141 &  25 & 17.7\% \\
\midrule
\textbf{Total}             & \textbf{1{,}638} & \textbf{504} & \textbf{30.8\%} \\
\bottomrule
\end{tabular}
\caption{Per-repository acceptance rates for SFT data generated by \polar with
Qwen3.5-122B-A10B and the \texttt{pi} harness on SWE-Gym. ``Accepted'' means
the agent's patch passed both \texttt{FAIL\_TO\_PASS} and \texttt{PASS\_TO\_PASS}
tests in the SWE-Bench evaluator.}
\label{tab:swegym-pi-rates}
\end{table}

\paragraph{Released format.}
Each accepted row contains the SWE-Gym instance metadata
(\texttt{instance\_id}, \texttt{repo}, \texttt{problem\_statement},
\texttt{base\_commit}, \texttt{version}) and the full multi-turn conversation
as a list of OpenAI-style messages with \texttt{role}, \texttt{content},
\texttt{tool\_calls}, and \texttt{tool\_call\_id} fields, terminated by the
assistant turn that produced the accepted patch. Trajectories are long: an
average of 104 messages per session and 51 assistant turns, with a long tail
above 200 turns. The corpus is released as a HuggingFace dataset under an
Apache-2.0 license, with a 90/10 train/test split stratified by repository so
that every repo is represented in both splits.\footnote{Available at
\url{https://huggingface.co/datasets/nvidia/polar-swegym-pi-qwen35-122b-a10b-trajectories}.}

We deliberately kept the filter narrow---a single binary verifier from the
existing SWE-Bench harness---to keep the case study reproducible. The same
\polar deployment can be re-used for richer offline pipelines without changing
the runtime: rejection sampling falls out of running multiple completions per
prompt and keeping only those that pass the verifier; verifier-training data
falls out of retaining the rejected trajectories alongside the accepted ones;
preference data falls out of pairing accepted and rejected traces from the
same prompt. Scaling the present run to the full 2{,}438-instance SWE-Gym set,
swapping in stronger teachers, or adding additional harnesses (e.g.\
\texttt{codex} or \texttt{claude\_code}) requires no changes to the
orchestration code---only additional submitter shards and the corresponding
checkpoint.

\section{Conclusion}
\polar treats agent test-time environments as a first-class part of the RL system rather than an implementation detail to be ported into the trainer. Its central design choice is to move the integration boundary to the model endpoint: the harness runs normally, the proxy observes token-level model traffic, and the rollout service turns completed executions into trainable trajectories and rewards. This separation lets rollout scale independently from training and inference, while preserving the behavior of non-standard harnesses whose value often lies in their engineering details. We believe \polar opens a new paradigm for scaling agentic RL infrastructure in the modern era, and we are actively developing and maintaining the framework as the ecosystem evolves.

\newpage
\bibliographystyle{plainnat}
\bibliography{reference}

\clearpage
\newpage

\newpage
\appendix
\onecolumn
\section{Appendix}

\subsection{Framework Comparison}
\label{app:framework_comparison}

\begin{table}[H]
\centering
\scriptsize
\setlength{\tabcolsep}{4.5pt}
\renewcommand{\arraystretch}{1.15}
\begin{tabular}{lcccc}
\toprule
\textbf{System} &
\makecell{\textbf{Async RL}\\\textbf{Support}} &
\makecell{\textbf{Async}\\\textbf{Rollout Staging}} &
\makecell{\textbf{Rollout as}\\\textbf{Service}} &
\makecell{\textbf{Agent Harness}\\\textbf{Agnostic}} \\
\midrule
\polar & \cmark & \cmark & \cmark & \cmark \\
\prorlagent \citep{prorlagent2026} & \cmark & \cmark & \cmark & \xmark \\
SkyRL-Agent \citep{skyrlagent2025} & \cmark & \cmark & \xmark & \pmark \\
PRIME-RL \citep{primerl2026} & \cmark & \xmark & \xmark & \xmark \\
Agent Lightning \citep{agentlightning2025} & \pmark & \xmark & \pmark & \pmark \\
rLLM \citep{rllm2025} & \pmark & \xmark & \xmark & \xmark \\
OpenClaw-RL \citep{openclawrl2026} & \cmark & \xmark & \xmark & \pmark \\
\bottomrule
\end{tabular}
\caption{\textbf{Comparing rollout-system design choices.} We find that modern rollout infrastructures should meet following criteria: async RL support means training can consume rollouts while generation continues under explicit policy-version or staleness handling; async rollout staging means rollout execution is decomposed into independently scheduled runtime-preparation, execution, post-run reconstruction/evaluation, and cleanup stages; rollout as service means a durable task API that is separable from a specific trainer loop; and native-harness agnosticism means a CLI, SDK, or application harness can be trained without being reimplemented as the framework's environment. \cmark{} denotes first-class support, \pmark{} denotes partial or planned support, and \xmark{} denotes that we did not find the property as a primary design contract in the referenced code or documentation.}
\label{tab:framework_comparison}
\end{table}

The partial marks in \cref{tab:framework_comparison} avoid treating adjacent mechanisms as absent. SkyRL exposes custom generators and Harbor integration, but native harnesses are not the default unit of rollout. Agent Lightning provides a rollout store, queue, runner control plane, and broad framework instrumentation, while its main boundary is trace/workflow observability rather than staged execution of opaque harness processes. rLLM includes fully asynchronous training and a model gateway that captures token IDs and log probabilities, but its rollout service abstraction is narrower than a distributed runtime lifecycle service. OpenClaw-RL decouples serving, rollout collection, judging, and training for real-world agent settings; its support is organized around OpenClaw and specific terminal, GUI, SWE, and tool-call recipes rather than arbitrary native harness submission.

\subsection{SWE-Gym GRPO Hyperparameters}
\label{app:swegym_grpo_hparams}

\begin{table}[H]
\centering
\small
\setlength{\tabcolsep}{6pt}
\renewcommand{\arraystretch}{1.12}
\begin{tabular}{ll}
\toprule
\textbf{Hyperparameter} & \textbf{Value} \\
\midrule
Base checkpoint & \texttt{Qwen/Qwen3.5-4B} \\
Training data & \texttt{NovaSky-AI/SkyRL-v0-293-data}, train split, 293 tasks \\
Trainer & Slime asynchronous GRPO \\
Epochs & 1 \\
Rollout batch size & 4 \\
Samples per prompt & 16 \\
Trace construction & \texttt{prefix\_merging} \\
Optimizer & Adam \\
Learning rate & $1\times10^{-6}$ \\
Weight decay & 0.1 \\
TIS & Enabled \\
\bottomrule
\end{tabular}
\caption{\textbf{Training hyperparameters for the SWE-Gym GRPO experiments.} The table reports ordinary policy-optimization and rollout parameters from \texttt{examples/swegym\_slime\_grpo}; cluster topology and worker placement are omitted.}
\label{tab:swegym_grpo_hparams}
\end{table}

\subsection{Representative Task Payload}
\label{app:task_payload}

\begin{polarlisting}{Representative Polar Task Payload}
{
  "task_id": "polar-swegym-{rollout_id}-{group_index}",
  "instruction": "Fix the issue in /polar/session/workspace.",
  "num_samples": 8,
  "timeout_seconds": 1200,
  "runtime": {
    "backend": "docker",
    "image": "{sample.metadata.runtime_image}",
    "network": "host",
    "workdir": "/polar/session/workspace",
    "prepare": [
      {
        "type": "exec",
        "command": "prepare repository, harness, and dependencies"
      }
    ]
  },
  "agent": {
    "harness": "codex",
    "model_name": "{served_model_name}"
  },
  "builder": {
    "strategy": "prefix_merging"
  },
  "evaluator": {
    "strategy": "swebench_harness",
    "refresh_runtime": true,
    "config": {
      "repo_dir": "/testbed",
      "patch_command": "cd /polar/session/workspace && git add -A && git diff --cached --binary",
      "instance": "{sample.metadata.instance}"
    }
  },
  "callback_url": "http://{trainer_host}:{callback_port}/callback/task_result",
  "metadata": {
    "group_id": "{rollout_group_id}",
    "policy_version": "{policy_version}",
    "rollout_step": "{rollout_step}"
  }
}
\end{polarlisting}

\subsection{Representative Trace}
\label{app:trace_example}

\begin{polarlisting}{Representative Trainer-Facing Trace}
{
  "prompt_ids": [151644, 872, "..."],
  "response_ids": [9211, 374, 264, "...", 151645, 271, 151644],
  "loss_mask": [1, 1, 1, "...", 0, 0, 1],
  "response_logprobs": [
    {
      "token": "The",
      "token_id": 785,
      "logprob": -0.21
    },
    {
      "token": " patch",
      "token_id": 10042,
      "logprob": -0.44
    },
    {
      "token_id": 151645,
      "logprob": 0.0
    }
  ],
  "prompt_messages": [
    {"role": "system", "content": "agent harness system prompt"},
    {"role": "user", "content": "task instruction and current tool state"}
  ],
  "response_messages": [
    {"role": "assistant", "content": "sampled assistant turn"}
  ],
  "tools": null,
  "finish_reason": "stop",
  "reward": 1.0,
  "metadata": {
    "session_id": "session-123",
    "task_id": "polar-swegym-0001",
    "builder": "prefix_merging",
    "harness": "codex"
  }
}
\end{polarlisting}

\subsection{Service API Summary}

The rollout service exposes a small asynchronous API:

\begin{itemize}
    \item \texttt{POST /rollout/task/submit}: submit a non-blocking task request.
    \item \texttt{GET /rollout/task/\{task\_id\}}: poll task status, partial results, and final results.
    \item \texttt{GET /rollout/status}: inspect task states, node states, and pending sessions.
    \item \texttt{POST /callbacks/session\_result}: receive gateway session callbacks.
    \item \texttt{POST /nodes/register} and \texttt{POST /nodes/\{node\_id\}/heartbeat}: maintain gateway membership and scheduling metrics.
\end{itemize}

The gateway exposes a control surface for session creation, status, and deletion, plus a catch-all proxy surface for provider-style model requests. Session deletion is used by the rollout pipeline as best-effort cleanup after a terminal result has been persisted.


\end{document}